\documentstyle[aps,prc,psfig]{revtex}
\begin{document}
\draft
\title{Mass formula for $T=0$ and $T=1$ ground states of $N=Z$ nuclei.}
\author{E.~Baldini-Neto and C.L.~Lima}
\address{Nuclear Theory and Elementary Particle Phenomenology Group,\\
Instituto de F\'\i sica, Universidade de S\~ao Paulo,\\
C.P. 66318, 05315-970 S\~ao Paulo, S\~ao Paulo, Brazil}
\author{P.~Van Isacker}
\address{Grand Acc\'el\'erateur National d'Ions Lourds, \\
BP 55027, F-14076 Caen Cedex 5, France}
\date{\today }
\maketitle

\begin{abstract}
An algebraic model is developed
to calculate the $T=0$ and $T=1$
ground-state binding energies for $N=Z$ nuclei.
The method is tested in the $sd$ shell
and is then extended to the 28--50 shell
which is currently the object of many experimental studies.
\end{abstract}

\newpage
\section{Introduction}
\noindent
The investigation of the behavior of nuclei under extreme conditions
has become an important tool to reveal new facets of nuclear matter.
In particular, nuclei at the frontiers of the valley of stability
constitute nowadays the most active research area
of nuclear structure physics.
With the advent of new radioactive beam facilities
it is now possible to produce exotic nuclei
that may have occurred naturally
in the interior of exploding supernovas~\cite{RevExp}.
In short, extremely proton- and neutron-rich nuclei
are now within experimental reach.
Of specific interest to the present paper
are the considerable experimental efforts
to study nuclei with roughly the same number of neutrons and protons,
$N\approx Z$.

On the theoretical side,
the challenge is to investigate whether models,
developed for---and using the phenomenology of---stable nuclei,
can still be applied in these new, as yet uncharted regions
and, if not, to propose new approaches to do so
based on the data available up to now.
One of the main open questions is the validity
of the nuclear shell model
with its traditional magic numbers
and of the usual treatment of the residual interaction 
among the valence nucleons~\cite{nois,np}.

The nuclear mass is a property of quintessential importance
as it directly determines the stability of a nucleus.
There are several theoretical approaches
that reproduce the systematics of masses of nuclei
and it is worthwhile to mention here two of them.
The Extended Thomas-Fermi plus Strutinsky Integral~\cite{To 84} (ETFSI)
is a high-speed approximation to the Hartree-Fock (HF) method
with pairing correlations taken into account through BCS theory.
In earlier versions a Wigner term was not included
and this has been claimed to be the reason
for the systematic calculated underbinding by about 2 MeV
for even-even $N=Z$ nuclei~\cite{Ab 92}.
This effect persists for $N=Z$ odd-odd systems
and for $N=Z\pm1$ odd-mass nuclei but with less prominence.
The mass formula
based on the Finite Range Droplet Model (FRDM)~\cite{Ni 95}
starts from a sophisticated liquid drop mass formula
to which microscopic corrections due to shell effects are added.
Both approaches have comparable numbers of parameters (about 15)
and make reliable predictions with impressive success.
In the FRDM and also in a recent ETFSI calculation~\cite{Go 01} 
a Wigner (correction) term is included
that specifically deals with the peculiar behavior
of binding energies of $N\approx Z$ nuclei
and has a cusp-like behavior for $N=Z$.
This treatment is effective for known masses
but, as the correction is entirely {\it ad hoc},
it has the drawback
that an extrapolation to unknown nuclei can be dangerous. 
It is therefore of interest to develop models
based on simple physical principles
that can account for the behavior of nuclear masses
at the $N\approx Z$ line.

Many models have been used over the past years
to investigate the structure of heavier $N\approx Z$. 
We mention in particular recent applications
of the Hartee-Fock-Bogolyubov (HFB) method
that includes proton-neutron pairing correlations~\cite{Go 99}.
This approach is tailor-made
for the treatment of $N\approx Z$ nuclei
but has the drawback of the lack of particle-number projection.
Shell-model calculations~\cite{Ap 98}
are generally extremely successful
in reproducing spectroscopic nuclear data
but require large configuration space diagonalizations.
In addition, no reliable theoretical procedure exists
for deriving the monopole part of the residual interaction
which hinders a comparison with measured nuclear binding energies.
An algebraic approach~\cite{Lp 97},
which has affinities with the one presented here,
utilizes the concept of dynamical supersymmetry
for the calculation of the binding energies
in the $sd$ shell but does not go beyond it.

In this paper,
the Interacting Boson Model (IBM)~\cite{Ia 87}
in its isospin invariant version
is applied to proton-rich $N\approx Z$ nuclei.
The goal is to obtain reliable estimates of binding energies
of $T=0$ and $T=1$ ground states in self-conjugate ($N=Z$) nuclei
based on the concept of dynamical symmetry.
The mass formula proposed is relatively simple
and contains terms with an intuitively understandable significance.
A particular ingredient
is its treatment of the competition
between isoscalar and isovector pairing.

\section{An IBM-4 mass formula}
\noindent
The Interacting Boson Model in its original version (IBM-1)~\cite{Ar 75}
successfully describes collective aspects of nuclei 
through the use of $s$ and $d$ bosons
which are thought to approximate pairs of valence nucleons
coupled to angular momentum 0 and 2.
No distinction is made between neutron and proton bosons.
Whenever the difference between the neutron and proton fluids
is thought to play a role,
one is forced to use more elaborate versions of the IBM.
The neutron-proton interacting boson model, or IBM-2,
was introduced mainly to provide
a microscopic foundation to the model~\cite{Ot 78}.
It uses as building blocks $s$ and $d$ bosons
constructed from neutron-neutron (nn) and proton-proton (pp) pairs solely.
In the third and fourth versions of IBM, IBM-3 and IBM-4,
the isospin quantum number is introduced in a natural way.
In IBM-3 the entire isospin triplet $T=1$ is included,
leading to nn, np, and pp pairs with $T_z=+1,0,-1$~\cite{Ew 80}.
The IBM-4 considers both $T=0$ and $T=1$ pairs;
the $T=1$ bosons are assigned an intrinsic spin $S=0$
while $T=0$ bosons carry an intrinsic spin $S=1$~\cite{Ee 81}. 
A justification of this choice
is that the two-particle isospin-spin combinations
$(TS)=(10)$ and $(TS)=(01)$
are lowest in energy
and that they give rise to an SU(4) algebra
which is the boson equivalent of Wigner's supermultiplet algebra~\cite{Wg 37}. 

The mass region $28\leq N\approx Z\leq 50$
has a very rich structural behavior,
presenting many aspects of nuclear motion.
It is an ideal testing ground for various models
since a proper description of the data
relies on the interplay between $T=0$ as well as $T=1$ pairing
and deformation-driving interactions.
In addition, it is a region of intense experimental studies 
but with few experimental data available up to now.

Very recently, the IBM-4 was applied
to the spectroscopy of exotic $N\approx Z$ nuclei
in the $pf_{5/2}g_{9/2}$ shell~\cite{Oj 01}.
In this approach the IBM-4 Hamiltonian is derived
from a realistic shell-model Hamiltonian
through a mapping carried out for $A=58$ and 60 nuclei.
The boson energies and the boson-boson interactions
are thus derived microscopically
and no parameter enters the calculation
(since the shell-model interaction is considered as an input).
This microscopically derived Hamiltonian
gives good results in $^{62}$Ga
(when compared to the shell model)
and predicts the energy spectra of heavier $N=Z$ nuclei
(such $^{66}$As and $^{70}$Br).
The approach is reasonably successful
in obtaining a spectroscopy of low-spin states in $N\approx Z$ nuclei.
It makes use, however, of a complicated Hamiltonian
and, moreover, calculations beyond $^{70}$Br seem difficult.

Prompted by these considerations,
in particular the need for reliable binding energy predictions
at the $N=Z$ line
and the existence of a microscopically derived IBM-4 Hamiltonian,
we propose here a simple mass formula in the context of IBM-4.
In previous work~\cite{Vi 97}
one of us introduced an algebraic Hamiltonian
(which can be regarded as the $s$-boson channel
of the general IBM-4 Hamiltonian of Ref.~\cite{Oj 01})
with the specific aim
to study the competition
between the isovector and isoscalar pairing modes
in self-conjugate nuclei.
The $s$ bosons do give rise only
to the symmetric representation of ${\rm U}_L(6)$
but this approximation can be justified
in even-even and odd-odd $N=Z$ nuclei
(the only ones considered here)
where this is indeed the favored U(6) representation~\cite{Ee 81}.

These previous studies~\cite{Oj 01,Vi 97} suggest
that the relevant terms in a simple IBM-4 mass formula
must be taken from two different symmetry classifications:
\begin{eqnarray}
&&{\rm U}(36)\supset
\left({\rm U}_L(6)\supset\cdots\supset{\rm SO}_L(3)\right)
\nonumber\\&&\qquad\qquad\otimes
\left({\rm U}_{TS}(6)\supset
\left\{\begin{array}{c}
{\rm SU}_{TS}(4)\\
{\rm U}_T(3)\otimes{\rm U}_S(3)
\end{array}\right\}
\supset{\rm SO}_T(3)\otimes{\rm SO}_S(3)
\right).
\label{chain}
\end{eqnarray}
Due to the overall symmetry of the boson state under U(36),
the ${\rm U}_L(6)$ and ${\rm U}_{TS}(6)$ representations
have identical labels $[N_1,\dots,N_6]$.

The terms chosen from~(\ref{chain})
for the calculation of binding energies of $N=Z$ nuclei
are the following.
First, the linear and quadratic Casimir operators of ${\rm U}_L(6)$
are included.
The total number of bosons, $N$,
labels the symmetric representations of ${\rm U}_L(6)$
and, as a result, these Casimir operators
take account of the bulk properties of the nucleus
and lead to a smooth variation of the mass with particle number.
The next two terms to be included
are the quadratic Casimir operator of ${\rm SU}_{TS}(4)$
and the linear Casimir operator of ${\rm U}_S(3)$.
In the $L=0$ channel considered here,
these include the effects of $T=0$ and $T=1$ pairing interaction;
${\rm SU}_{TS}(4)$ implies equal $T=0$ and $T=1$ strengths
while ${\rm U}_S(3)$ breaks this degeneracy
by removing all non-scalar ${\rm U}_S(3)$ states
from the low-energy spectrum.
In Ref.~\cite{Vi 97}
the {\em quadratic} Casimir operator of ${\rm SU}_S(3)$
was considered;
here the {\em linear} Casimir operator of ${\rm U}_S(3)$ is preferred
because it is shown since to be connected
with the spin-orbit term in the mean field
which breaks the isoscalar-isovector degeneracy
in favor of isovector states~\cite{Oj 00}. 
The final term considered
is the quadratic Casimir operator of ${\rm SO}_T(3)$, $T(T+1)$,
which is known to represent
the nuclear symmetry and Wigner energies.
In summary, the following Hamiltonian is taken:
\begin{equation}
\hat H=
E_0
+\alpha\hat C_1[{\rm U}_L(6)]
+\beta\hat C_2[{\rm U}_L(6)]
+\gamma\hat C_2[{\rm SU}_{TS}(4)]
+\xi\hat C_1[{\rm U}_S(3)]
+\eta\hat C_2[{\rm SO}_T(3)],
\label{hamil}
\end{equation}
where $E_0$ is the energy of the doubly magic core,
specific for a given mass region.
Note the omission from~(\ref{hamil}) of operators
associated with ${\rm U}_{TS}(6)$, ${\rm U}_T(3)$, and ${\rm SO}_S(3)$;
these are not needed because their effect is equivalent
to the corresponding operators of
${\rm U}_L(6)$, ${\rm U}_S(3)$, and ${\rm SO}_T(3)$.

All operators in~(\ref{hamil}) mutually commute,
except for $\hat C_2[{\rm SU}_{TS}(4)]$
and $\hat C_1[{\rm U}_S(3)]$
and hence the solution of $\hat H$
involves a numerical diagonalisation
which is most conveniently done in the second basis in~(\ref{chain}),
labeled as $|[N](\lambda_T)T\otimes(\lambda_S)S\rangle$.
The matrix elements of $\hat C_2[{\rm SU}_{TS}(4)]$
can be calculated analytically~\cite{Vi 97},
\begin{eqnarray}
V_{\lambda_T\lambda_S\lambda_T^\prime\lambda_S^\prime}^{NTS}
&\equiv&
\langle[N](\lambda_T)T\otimes(\lambda_S)S|
\hat C_2[{\rm SU}_{ST}(4)]
|[N](\lambda_T^\prime)T\otimes(\lambda_S^\prime)S\rangle,
\nonumber\\
V_{\lambda_T\lambda_S\lambda_T\lambda_S}^{NTS}
&=&
2\lambda_T\lambda_S+3N+T(T+1)+S(S+1),
\nonumber\\
V_{\lambda_T\lambda_S\lambda_T-2\lambda_S+2}^{NTS}
&=&
[(\lambda_T-T)(\lambda_T+T+1)(\lambda_S-S+2)(\lambda_S+S+3)]^{1/2}, 
\nonumber\\
V_{\lambda_T\lambda_S\lambda_T+2\lambda_S-2}^{NTS}
&=&
[(\lambda_T-T+2)(\lambda_T+T+3)(\lambda_S-S)(\lambda_S+S+1)]^{1/2},
\label{matel}
\end{eqnarray}
while the other operators are diagonal with eigenvalues given by
\begin{eqnarray}
\langle[N](\lambda_T)T\otimes(\lambda_S)S|
\hat C_1[{\rm U}_L(6)]
|[N](\lambda_T)T\otimes(\lambda_S)S\rangle
&=&N,
\nonumber\\
\langle[N](\lambda_T)T\otimes(\lambda_S)S|
\hat C_2[{\rm U}_L(6)]
|[N](\lambda_T)T\otimes(\lambda_S)S\rangle
&=&N(N+5),
\nonumber\\
\langle[N](\lambda_T)T\otimes(\lambda_S)S|
\hat C_1[{\rm U}_S(3)]
|[N](\lambda_T)T\otimes(\lambda_S)S\rangle
&=&\lambda_S,
\nonumber\\
\langle[N](\lambda_T)T\otimes(\lambda_S)S|
\hat C_2[{\rm SO}_T(3)]
|[N](\lambda_T)T\otimes(\lambda_S)S\rangle
&=&T(T+1).
\label{matdia}
\end{eqnarray}

\section{Results}
\begin{figure}
\centerline{\psfig{figure=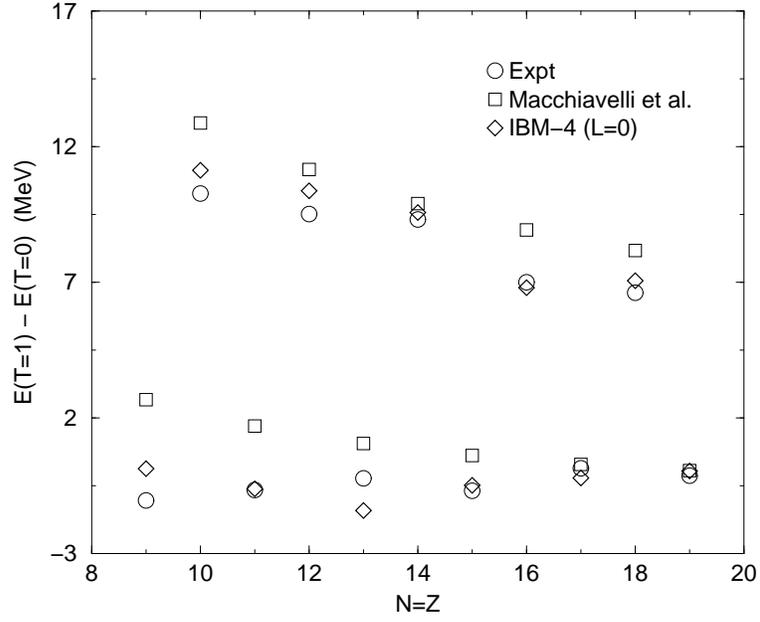,width=10.0cm}}
\caption{Calculated energy differences $E(T=1)-E(T=0)$
in $N=Z$ $sd$-shell nuclei
for the parameters given in Table~\protect\ref{parameters},
compared with the experimental differences
and those of Macchiavelli {\it et al.}~\protect\cite{Ma 00}.}
\label{sd}
\end{figure}

\noindent
We first apply the mass formula to $N=Z$ nuclei in the $sd$ shell,
from $^{16}$O to $^{40}$Ca,
where the experimental masses are well known~\cite{Aw 95}.
The five parameters of the Hamiltonian~(\ref{hamil})
are adjusted to the binding energies
of the lowest states with $T=0$ and those with $T=1$
of all even-even and odd-odd self-conjugate $sd$-shell nuclei.
There are thus two data points per nucleus,
which is crucial for a reliable determination
of the parameters $\gamma$, $\xi$, and $\eta$. 
Binding energies are corrected for Coulomb effects
according to the prescription given in~\cite{Ni 95}.
A drawback of the present formula
is the occurrence of a discontinuity at mid-shell
which is related to a change of core (from $^{16}$O to $^{40}$Ca).
To avoid these mid-shells effects,
two different fits are performed for each half of the shell,
a first one for nuclei from $^{18}$F up to $^{28}$Si ($N=6$ bosons)
and a second one (with $^{40}$Ca as a core)
for nuclei from $^{38}$K down to $^{30}$P. 
In Table~\ref{parameters} the two parameter sets are given
in the lines labeled `$^{16}$O to $^{28}$Si'
and `$^{30}$P to $^{40}$Ca'.
The major difference between the two sets
is the sign change in $\alpha$
which is required since in the first half
$N$ counts the pairs of nucleons {\em added} to $^{16}$O
while in the second half
it counts the pairs {\em subtracted} from $^{40}$Ca.
One also notes that
$\alpha$ has a larger absolute value in first half than in the second:
this must be so since, in the $sd$ shell,
the binding energy per nucleon increases
as the size of the nucleus grows.
Furthermore, the parameters $\gamma$, $\xi$, and $\eta$
decrease (in absolute value),
as a result of the average interaction strength
which decreases with mass.
Nevertheless, one notes that this decrease
is stronger for $\gamma$ than it is for $\xi$,
that is, the ratio $|\xi/\gamma|$ is larger
in the second half of the $sd$ shell than it is in the first.
Again, this is understandable intuitively
because one expects the Wigner SU(4) symmetry
to be increasingly broken by the spin-orbit term [${\rm U}_S(3)$].
The resulting binding energies for each half of the shell
are shown in Tables~\ref{sd1} and~\ref{sd2}, respectively.
Also the isospin of each state is indicated
as well as the difference $\Delta E$
between the calculated and measured binding energies.
In Fig.~\ref{sd} the differences in energy
between the $T=1$ and $T=0$ states 
are compared with the observed ones
and also with the semi-empirical formula for this quantity
given in Ref.~\cite{Ma 00}. 
The root-mean-squared (rms) deviation is 0.876 MeV
in the first half of the $sd$ shell
and 0.245 MeV in the second half.
Since reasonable results are obtained
with parameters that can be qualitatively understood
from simple arguments,
an extension towards the 28--50 shell can be considered.

\begin{figure}
\centerline{\psfig{figure=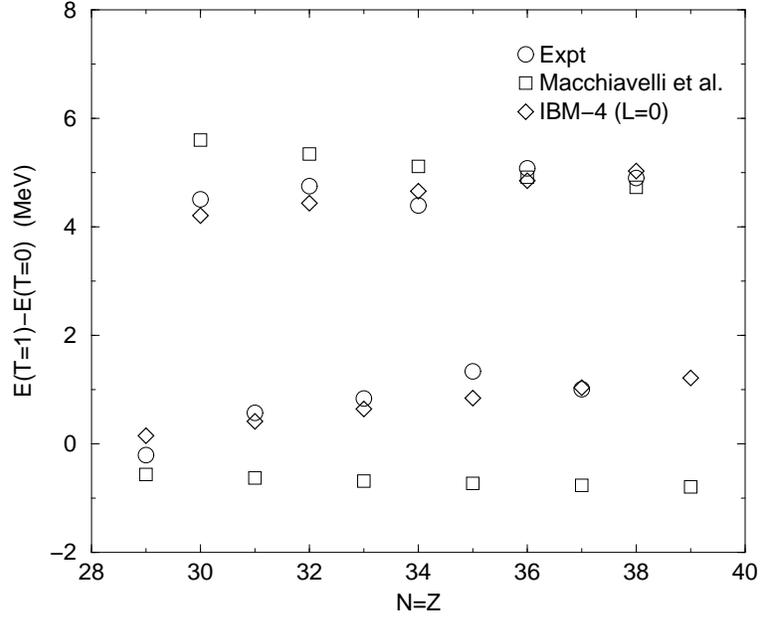,width=10.0cm}}
\caption{Calculated energy differences $E(T=1)-E(T=0)$
in $N=Z$ nuclei between $^{58}$Cu to $^{78}$Y
for the parameters given in Table~\protect\ref{parameters},
compared with the experimental differences
and those of Macchiavelli {\it et al.}~\protect\cite{Ma 00}.}
\label{pf1}
\end{figure}

\begin{figure}
\centerline{\psfig{figure=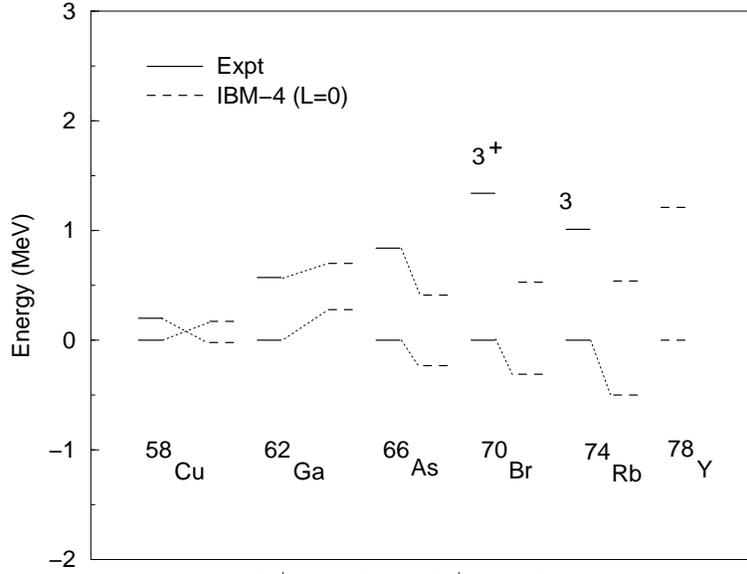,width=10.0cm}}
\caption{Experimental and calculated energies
of $(1^+,T=0)$ and $(0^+,T=1)$ levels
in odd-odd $N=Z$ from $^{58}$Cu to $^{78}$Y.
In $^{70}$Br, $^{74}$Rb, and $^{78}$Y
the $(1^+,T=0)$ levels are not known experimentally
and in the former two nuclei the angular momentum
of the lowest (known) $T=0$ state is indicated.}
\label{pfoo}
\end{figure}

We begin with a discussion of 
the first half of the 28--50 shell,
for nuclei ranging from $^{58}$Cu to $^{78}$Y.
The ground state of all these self-conjugate nuclei is a $0^+$,
with either $T=0$ in even-even or $T=1$ in odd-odd nuclei,
with the exception of $^{58}$Cu
which has a $(J^\pi,T)=(1^+,0)$ ground state.
Up to $^{64}$Ge the masses are well known
and can be taken from the compilation of Audi and Wapstra~\cite{Aw 95}.
Of the heavier $N=Z$ nuclei,
also the masses of $^{72}$Kr and $^{74}$Rb
are listed by Audi and Wapstra.
The masses of $^{66}$As and $^{68}$Se
are available from a recent measurement~\cite{Al 01}
and that of $^{76}$Sr from Ref.~\cite{La 01}.
The latter experiment also gives a mass for $^{68}$Se
but since it is far off the systematics of Audi and Wapstra,
the result from~\cite{Al 01} is used.
The mass of $^{70}$Br is not known experimentally
but as it in the middle of a region of nuclei
with measured masses close to the extrapolations
of Audi and Wapstra,
we have adopted their extrapolated value for $^{70}$Br.
The mass of $^{78}$Y is not known and not included in the fit.
The binding energies of the lowest $T=1$ states
in even-even $N=Z$ nuclei
are derived from those of the isobaric analogues
(also taken from Ref.~\cite{Aw 95})
after an appropriate Coulomb correction.
The evolution of the splitting between
$(J^\pi,T)=(0^+,1)$ and $(J^\pi,T)=(1^+,0)$ states
in odd-odd nuclei is of particular interest
as regards the question of $T=0$ and $T=1$ pairing
and is currently the object of several experimental studies.
The $(0^+,1)$ state in $^{58}$Cu
lies 0.202 MeV above the $(1^+,0)$ ground state~\cite{Nd 99}.
This order is reversed in $^{62}$Ga
where the $(1^+,0)$ state
is 0.571 MeV above the $(0^+,1)$ ground state~\cite{Vi 98}.
The $E(0^+,1)-E(1^+,0)$ splitting then continues to rise
to 0.837 MeV in $^{66}$As~\cite{Gr 98}.
A very recent experiment on $^{70}$Br~\cite{Je 01}
has not observed a $(1^+,0)$ level;
the lowest observed $T=0$ level (with $J^\pi=3^+$)
is at an excitation energy of 1.337 MeV.
Similarly, the lowest $T=0$ state in $^{74}$Rb
measured by Rudolph {\it et al.}~\cite{Ru 96}
at an excitation energy of 1.006 MeV has $J=3$
and the energy of the $1^+$ state is unknown.
With these data as input,
the parameters in~(\ref{hamil}) can be adjusted
through a fit procedure
that minimizes the rms deviation
in the binding energies of two states per nucleus (if known).
The resulting parameters
are shown in the line labeled `$^{56}$Ni to $^{78}$Y'
of Table~\ref{parameters}
and lead to an rms deviation of 0.396 MeV.
In Fig.~\ref{pf1} the differences in energy
between the $T=1$ and $T=0$ states 
are compared to the observed ones.
One notes the good agreement that is obtained
which gives confidence in the energy splittings
of 0.847, 1.037, and 1.214 MeV predicted in
$^{70}$Br, $^{74}$Rb, and $^{78}$Y, respectively.
As already mentioned, the energy difference $E(0^{+},1)-E(1^+,0)$
is not known experimentally in these isotopes.
In the former two, $^{70}$Br and $^{74}$Rb,
the energy difference with the lowest (known) $T=0$ state
is shown in Fig.~\ref{pf1}.
To emphasize the point that these energy splittings
result from a calculation of total binding energies,
the odd-odd results are represented
in a different way in Fig.~\ref{pfoo}.
Note that this plot implies a comparison
of {\em absolute} binding energies:
for representation purposes
the meausured binding energy of the ground state
of a particular nucleus
is drawn at zero and the other levels of that nucleus
are given relative to that ground-state energy.

\begin{figure}
\centerline{\psfig{figure=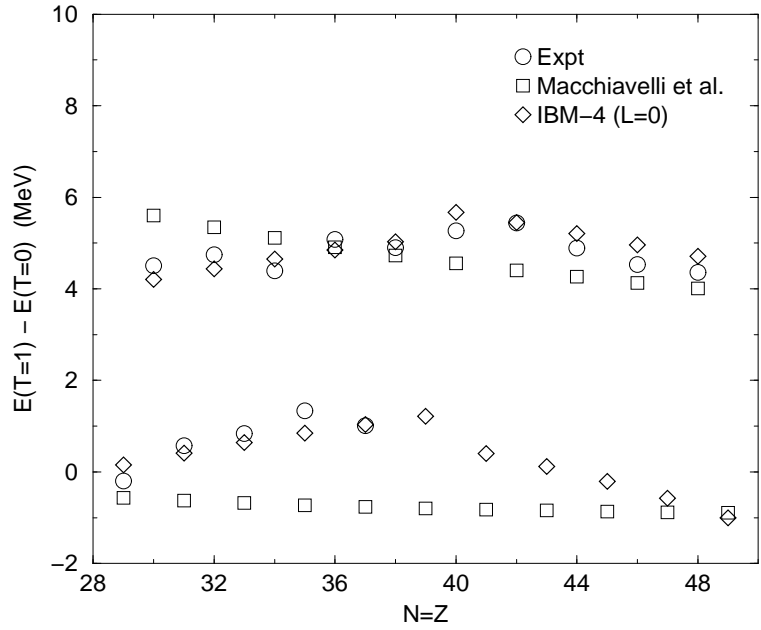,width=10.0cm}}
\caption{Energy differences $E(0^{+},1)-E(1^{+},0)$
for the entire 28--50 shell
with parameters fitted separately for each half (see text for details).
In the first half (up to $^{78}$Y) 
`Expt' refers to measured masses
while in the second half
it refers to the extrapolations of~\protect\cite{Aw 95}.
Also the results of Ref.~\protect\cite{Ma 00} are shown.}
\label{pf}
\end{figure}

For the second-half of the 28--50 shell
the situation is more complicated
since there are no data available.
The core is $^{100}$Sn with a ground-state mass
measured in Ref.~\cite{Ma 96}.
Since so little is known experimentally,
we use the extrapolations from Audi and Wapstra~\cite{Aw 95}
for the masses of even-even and odd-odd nuclei,
complemented with the results for $^{78}$Y
from the fit to the first half of the 28--50 shell.
The resulting parameters
are shown in the line labeled `$^{78}$Y to $^{100}$Sn'
of Table~\ref{parameters}.
The predictions for the splitting between $T=1$ and $T=0$ states
for the entire 28--50 shell 
are shown in Fig.~\ref{pf}.
One notes a satisfactory agreement with the data,
when available.
The use of extrapolated data, however,
should weaken the confidence in the predictions
for the $E(0^+,1)-E(1^+,0)$ splitting in odd-odd nuclei.

\section{Conclusions}
A simple mass formula based on IBM-4 has been proposed
to estimate the binding energies
of the lowest $T=0$ and $T=1$ states of self-conjugate nuclei.
It has linear and quadratic terms in the boson number
that account for the smooth variation of mass with particle number,
supplemented with three contributions
that have a clear physical meaning:
an SU(4), a spin-orbit and a $\hat T^2$ term.
It can be considered as a local mass formula
that gives predictions of a specific interest
to current experiments at the $N=Z$ line.
As an application we considered nuclei from $^{56}$Ni to $^{78}$Y
where predictions could be made for some of the heavier isotopes
currently under study.
Also the second half of the 28--50 shell was considered
although there predictions are more questionable
due to the lack of reliable data.

The advantage with respect to previous IBM-4 work~\cite{Oj 01}
is that the Hamiltonian used is much simpler
and that only the $L=0$ channel is considered.
The numerical diagonalization then becomes trivial
and the calculations can be performed, without much effort,
for arbitrary numbers of bosons.
This is much harder to achieve with the full version of IBM-4.
On the down side it should be noted
that this approach is restricted to $N=Z$ odd-odd nuclei
since other odd-odd nuclei have a dominant U(6) representation
which is non-symmetric
and which cannot be constructed from $s$ bosons only.
Also, deformation effects
which are present with $s$ {\em and} $d$ bosons
and which can be included through orbital operators
[associated with algebras represented by dots in~(\ref{chain})],
are outside the scope of the simple approach presented here.

\section*{Acknowledgments}
One of us (EBN) would like to thank FAPESP for financial support
and also the people from GANIL for the hospitality
shown during his stay where part of this work was carried out.

\newpage
\begin{center}
\begin{table}
\caption{Parameters (in MeV) for the 8--20 and 28--50 shells.}
\label{parameters}
\begin{tabular}{crrrrr}
Shell & $\alpha$ & $\beta$ & $\gamma$ & $\xi$ & $\eta$  \\
\hline
$^{16}$O to $^{28}$Si  & $ 16.060$ & $0.477$ & $ 0.190$ & $-6.146$ & $-3.009$ \\
$^{30}$P to $^{40}$Ca  & $-24.538$ & $0.110$ & $ 0.065$ & $-3.735$ & $-1.846$ \\
$^{56}$Ni to $^{78}$Y  & $ 22.815$ & $0.118$ & $-0.068$ & $-1.953$ & $-0.450$ \\
$^{78}$Y to $^{100}$Sn & $-28.464$ & $0.118$ & $-0.188$ & $-1.045$ & $-0.512$ \\
\end{tabular}
\end{table}

\begin{table}
\caption{Binding energies (in MeV) of $N=Z$ nuclei
in the first half of the $sd$ shell.
Calculated values are obtained
with the parameters given in Table~\protect\ref{parameters}.}
\label{sd1}
\begin{tabular}{crrrr}
Nucleus & $T$ & $E_{\rm Expt}$ & $E_{\rm IBM4}$ & $\Delta E$ \\ 
\hline
$^{18}$F  & 0 & $151.662$ & $152.573$ & $-0.912$ \\
$^{18}$F  & 1 & $150.620$ & $152.701$ & $-2.081$ \\
$^{20}$Ne & 0 & $178.307$ & $178.887$ & $-0.580$ \\
$^{20}$Ne & 1 & $168.033$ & $167.755$ & $ 0.278$ \\
$^{22}$Na & 0 & $195.476$ & $195.332$ & $ 0.143$ \\
$^{22}$Na & 1 & $194.819$ & $194.722$ & $ 0.097$ \\
$^{24}$Mg & 0 & $223.545$ & $222.918$ & $ 0.628$ \\
$^{24}$Mg & 1 & $214.029$ & $212.543$ & $ 1.486$ \\
$^{26}$Al & 0 & $241.423$ & $242.181$ & $-0.758$ \\
$^{26}$Al & 1 & $241.195$ & $240.774$ & $ 0.421$ \\
$^{28}$Si & 0 & $270.581$ & $271.029$ & $-0.448$ \\
$^{28}$Si & 1 & $261.265$ & $261.465$ & $-0.200$ \\ 
\end{tabular}
\end{table}

\begin{table}
\caption{Binding energies (in MeV) of $N=Z$ nuclei
in the second half of the $sd$ shell.
Calculated values are obtained
with the parameters given in Table~\protect\ref{parameters}.}
\label{sd2}
\begin{tabular}{crrrr}
Nucleus & $T$ & $E_{\rm Expt}$ & $E_{\rm IBM4}$ & $\Delta E$ \\ 
\hline
$^{30}$P  & 0 & $289.433$ & $289.456$ & $-0.024$ \\
$^{30}$P  & 1 & $288.756$ & $288.968$ & $-0.213$ \\
$^{32}$S  & 0 & $315.655$ & $315.300$ & $ 0.350$ \\
$^{32}$S  & 1 & $308.653$ & $308.507$ & $ 0.146$ \\
$^{34}$Cl & 1 & $334.744$ & $334.723$ & $ 0.021$ \\
$^{34}$Cl & 0 & $334.598$ & $334.938$ & $-0.340$ \\
$^{36}$Ar & 0 & $361.450$ & $361.513$ & $-0.063$ \\
$^{36}$Ar & 1 & $354.839$ & $354.456$ & $ 0.383$ \\
$^{38}$K  & 0 & $381.186$ & $381.351$ & $-0.165$ \\
$^{38}$K  & 1 & $381.056$ & $381.393$ & $-0.337$ \\ 
\end{tabular}
\end{table}
\end{center}

\newpage 

\end{document}